\documentclass[conference]{IEEEtran}
\IEEEoverridecommandlockouts
 \usepackage{braket}
\usepackage{cite}
\usepackage{amsmath,amssymb,amsfonts}
\usepackage{algorithmic}
\usepackage{graphicx}
\usepackage{textcomp}
\usepackage{xcolor}
\usepackage{subcaption}
\def\BibTeX{{\rm B\kern-.05em{\sc i\kern-.025em b}\kern-.08em
    T\kern-.1667em\lower.7ex\hbox{E}\kern-.125emX}}
\begin{document}

\makeatletter
\newcommand{\newlineauthors}{%
  \end{@IEEEauthorhalign}\hfill\mbox{}\par
  \mbox{}\hfill\begin{@IEEEauthorhalign}
}
\newcommand{\compactblockA}[1]{%
  {\small \begin{tabular}{@{}c@{}}#1\end{tabular}}%
}

\makeatother

\title{Quantum-Assisted Learning of Time-Dependent Parabolic PDEs\\
\thanks{The authors acknowledge the support of the Danish e-Infrastructure Consortium (DeiC) and the National Quantum Algorithm Academy (NQAA) through the Postdoctoral Scholarship 
under the project ``Quantum-Driven Solutions for Multi-Agent Systems and Advanced Computation''. This work was also partially supported by UID/00147- Research Center for Systems and Technologies (SYSTEC) - and the Associate Laboratory Advanced Production and Intelligent Systems (ARISE, 10.54499/LA/P/0112/2020) funded by Fundação para a Ciência e a Tecnologia, I.P./ MCTES through the national funds.}
}

\author{\IEEEauthorblockN{Nahid Binandeh Dehaghani}
\IEEEauthorblockA{\compactblockA{\textit{Dep. of Electronic Systems} \\
\textit{Aalborg University}\\
Aalborg, Denmark \\
nahidbd@es.aau.dk}}
\and
\IEEEauthorblockN{Ban Tran}
\IEEEauthorblockA{\compactblockA{\textit{Dept. of Computer Science} \\
\textit{Texas Tech University}\\
Lubbock, USA \\
bantran@ttu.edu}}
\newlineauthors
\newlineauthors
\IEEEauthorblockN{A. Pedro Aguiar}
\IEEEauthorblockA{\compactblockA{\textit{SYSTEC-ARISE} \\
\textit{Faculty of Engineering, University of Porto}\\
Porto, Portugal\\
pedro.aguiar@fe.up.pt}}
\and
\IEEEauthorblockN{Rafal Wisniewski}
\IEEEauthorblockA{\compactblockA{\textit{Dept. of Electronic Systems} \\
\textit{Aalborg University}\\
Aalborg, Denmark\\
raf@es.aau.dk}}
\and
\IEEEauthorblockN{Susan Mengel}
\IEEEauthorblockA{\compactblockA{\textit{Dept. of Computer Science} \\
\textit{Texas Tech University}\\
Lubbock, USA\\
susan.mengel@ttu.edu}}
}

\maketitle
\vspace{-0.4cm}
\begin{abstract}
We present a hybrid quantum-classical framework for solving general time-dependent parabolic partial differential equations (PDEs) using quantum variational circuits. 
Building on the QPINN approach, this method applies broadly to parabolic PDEs. To demonstrate its effectiveness, we focus on the 1D and 2D heat equations as representative examples and analyze its performance under constrained quantum resources. Our results show that the framework can accurately capture spatiotemporal dynamics, offering a promising direction for quantum-assisted scientific computing.
\end{abstract}


\section{Introduction}
Hybrid quantum-classical algorithms \cite{b3} are emerging as promising tools for solving scientific computing problems, especially those involving partial differential equations (PDEs). Recent studies have introduced hybrid architectures that integrate parameterized quantum circuits  with classical embedding networks \cite{b1}, such as the trainable embedding quantum physics-informed neural network (QPINN) \cite{b2}. This framework combines the expressive power of quantum circuits with the flexibility of classical embeddings, providing a scalable and NISQ-era-compatible solution for solving PDEs.
 
In this work, we adapt a hybrid quantum-classical framework to simulate 1D and 2D heat equations, serving as representative examples of spatiotemporal diffusion problems. We analyze the performance of the method under limited quantum resources and demonstrate that the hybrid model can accurately capture thermal dynamics with modest circuit depth and qubit counts. The overall architecture of the learning framework, combining classical embedding with quantum variational processing, is illustrated in Fig.~\ref{fig:architecture}.
\begin{figure}[htbp]
\centering
\includegraphics[scale=0.205]{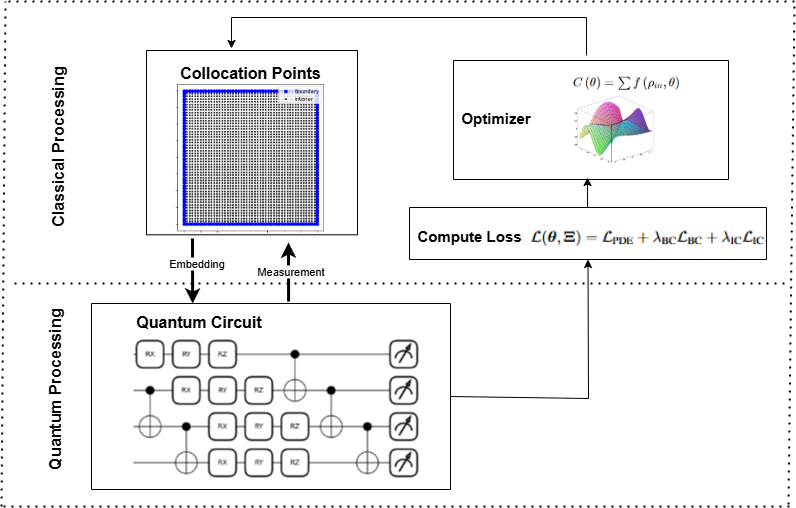}
\caption{
\small Classical embedding maps collocation points into angle parameters for a parameterized quantum circuit. The expectation values are used to compute the residual loss, which is minimized using a classical optimizer.}
\label{fig:architecture}
\vspace{-0.5cm}
\end{figure}

\vspace{-0.3cm}
\section{Theory and methodology}
We consider a general time-dependent parabolic partial differential equation (PDE) defined over a spatial domain \( \Omega \subset \mathbb{R}^d \) and a temporal interval \( t \in [0, T] \). 
 The residual forms of the governing equation, boundary conditions, and initial condition are given by:
\begin{align*}
\mathcal{D}(u(\mathbf{x}, t); \beta) - q(\mathbf{x}, t) &= 0, \quad (\mathbf{x}, t) \in \Omega \times (0, T], \\
\mathcal{B}(u(\mathbf{x}, t)) - b(\mathbf{x}, t) &= 0, \quad (\mathbf{x}, t) \in \partial\Omega \times (0, T], \\
u(\mathbf{x}, 0) - u_0(\mathbf{x}) &= 0, \quad \mathbf{x} \in \Omega.
\end{align*}
where \( \mathbf{x} \in \mathbb{R}^d \) denotes the spatial coordinate, \( t \in [0, T] \) is the time variable, \( u(\mathbf{x}, t) \) is the unknown solution to be approximated, \( \mathcal{D} \) is a differential operator (e.g., involving \( \partial u / \partial t \) and spatial derivatives) parameterized by physical constants \( \beta \), \( \mathcal{B} \) imposes boundary constraints, and
\( q(\mathbf{x}, t) \), \( b(\mathbf{x}, t) \), and \( u_0(\mathbf{x}) \) are known source, boundary, and initial functions.
Our objective is to construct an approximate solution \( \tilde{u}(\mathbf{x}, t) \) satisfying the PDE
over the full spatiotemporal domain by a quantum-assisted learning framework.

The quantum-assisted model predicts the solution \( \tilde{u}(\mathbf{x}, t; \boldsymbol{\theta}, \boldsymbol{\Xi}) \) by evaluating the expectation value of a quantum circuit defined as:\\
$\sum_{i=0}^{n} \left\langle 0 \middle| U_{\text{enc}}(\mathbf{x}, t; \boldsymbol{\Xi})^\dagger\, U_{\text{var}}(\boldsymbol{\theta})^\dagger\, O\, U_{\text{var}}(\boldsymbol{\theta})\, U_{\text{enc}}(\mathbf{x}, t; \boldsymbol{\Xi}) \middle| 0 \right\rangle,$
where \( n \) is the number of qubits in circuit, \( U_{\text{enc}}(\mathbf{x}, t; \boldsymbol{\Xi}) \) is the quantum encoding unitary mapping the classical input into a quantum state. The parameters \( \boldsymbol{\Xi} \) are weights of a classical neural network used to generate the angles for input encoding, \( U_{\text{var}}(\boldsymbol{\theta}) \) is a variational circuit block with trainable parameters \( \boldsymbol{\theta} \), \( \mathcal{O} \) is a measurement observable. The expectation value is evaluated over the initial quantum state \( \ket{0} \).

\paragraph*{Quantum Embedding Strategy}
To represent classical inputs \( (\mathbf{x}, t) \) as quantum states, we use a trainable embedding mechanism based on angle encoding. We construct a quantum feature map that rotates each qubit around the \( y \)-axis, where the rotation angle is computed using a classical neural network.

\paragraph*{Variational Circuit Design}
To capture complex solution dynamics, we construct a trainable quantum circuit \( U_{\text{var}}(\boldsymbol{\theta}) \), which acts as a feature transformation layer applied to the embedded quantum state. 
We adopt a layered architecture composed of single-qubit rotations and entangling operations that can be efficiently implemented on near-term hardware.

\paragraph*{Cost Function} The output of the variational quantum circuit is a scalar value computed via the expectation of a chosen observable \( \mathcal{O} \) applied to the transformed quantum state. We define the observable as a tensor product of Pauli-\( Z \) operators: $\mathcal{O} = \bigotimes_{i=1}^{n} {\sigma_z}_i$.
This observable measures the projection of each qubit in the computational basis and aggregates the outcome to estimate the solution field \( \tilde{u}(\mathbf{x}, t) \). 

\paragraph*{Loss Function}
We minimize a total loss function that combines PDE, initial condition, and boundary constraints, evaluated over collocation points sampled from the domain. \( \Omega \times (0,T] \), boundary \( \partial \Omega \times (0,T] \), and initial time \( t = 0 \).
The overall loss is expressed as
$\mathcal{L}(\boldsymbol{\theta}, \boldsymbol{\Xi}) = \mathcal{L}_{\text{PDE}} + \lambda_{\text{BC}} \mathcal{L}_{\text{BC}} + \lambda_{\text{IC}} \mathcal{L}_{\text{IC}}$,
where \( \lambda_{\text{BC}} \) and \( \lambda_{\text{IC}} \) are scalar weights controlling the emphasis on boundary and initial terms.
The individual loss components are defined as:\\
$\mathcal{L}_{\text{PDE}} = \sum_{(\mathbf{x}^j, t^j) \in \mathcal{S}_{\text{PDE}}} \left( \mathcal{D}(\tilde{u}(\mathbf{x}^j, t^j)) - q(\mathbf{x}^j, t^j) \right)^2,$ \\
$\mathcal{L}_{\text{BC}} = \sum_{(\mathbf{x}^j, t^j) \in \mathcal{S}_{\text{BC}}} \left( \mathcal{B}(\tilde{u}(\mathbf{x}^j, t^j)) - b(\mathbf{x}^j, t^j) \right)^2,$ \\
$\mathcal{L}_{\text{IC}} = \sum_{\mathbf{x}^j \in \mathcal{S}_{\text{IC}}} \left( \tilde{u}(\mathbf{x}^j, 0) - u_0(\mathbf{x}^j) \right)^2$.
The sets \( \mathcal{S}_{\text{PDE}}, \mathcal{S}_{\text{BC}}, \mathcal{S}_{\text{IC}} \) contain collocation points drawn from the interior, the boundary, and the initial surface of the spatiotemporal domain. 
\section{Results}
We begin by evaluating the framework on the 1D heat equation:
$\frac{\partial u(t,x)}{\partial t} = \alpha \frac{\partial^2 u(t,x)}{\partial x^2}$, $t \in [0, 0.95]$, $x \in [-1, 1]$, with initial condition \( u(0,x) = \sin(\pi x) \), where \( \alpha = \frac{0.01}{\pi} \) is the diffusion coefficient.
We discretize the domain with 100 spatial and 50 temporal collocation points. The model uses a 5-qubit, 4-layer variational ansatz and a classical FNN with two 10-neuron hidden layers and \texttt{tanh} activations, balancing expressivity with circuit depth. Fig.~\ref{fig:heat_prediction} illustrates the refernce solution, prediction and the absolute error. 
The results confirm that the model accurately approximates the true solution, with errors generally below \(10^{-3} \), primarily concentrated near boundaries.

\begin{figure}[htbp]
\centering
\includegraphics[scale=0.2]{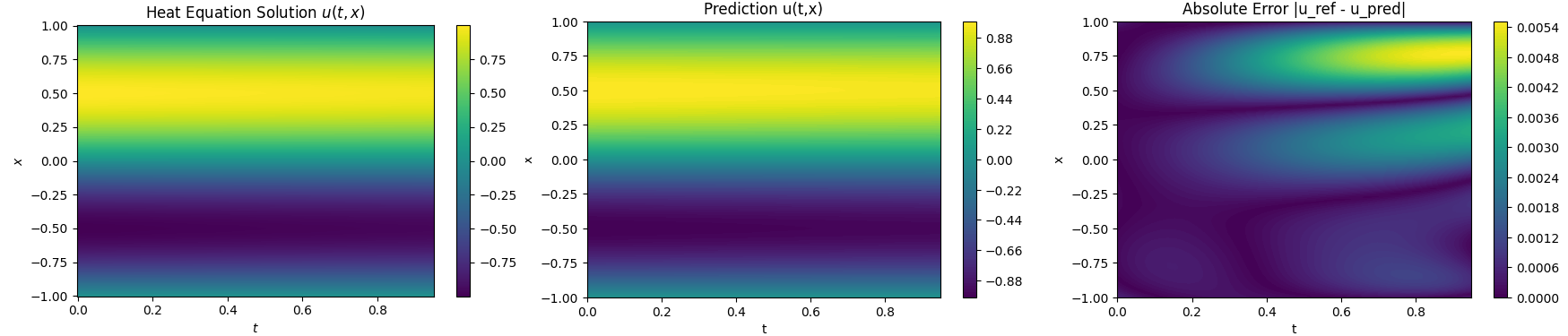} 
\caption{\small Reference solution, predicted solution and the absolute error for 1D heat equation.}
\label{fig:heat_prediction}
\vspace{-0.5cm}
\end{figure}

To validate the predictive performance of our QPINN-based framework on two-dimensional time-dependent problems, we simulate the 2D heat equation:
$\frac{\partial u(t, x, y)}{\partial t} = \kappa \left( \frac{\partial^2 u}{\partial x^2} + \frac{\partial^2 u}{\partial y^2} \right)$,
where \( \kappa = \frac{2}{\pi} \) is the thermal diffusivity constant.
The domain and initial condition are given by \( t \in [0, 0.1] \), spatial domain: \( (x, y) \in [-1, 1] \times [-1, 1] \), and initial condition $u(0, x, y) = \exp\left(-10(x^2 + y^2)\right)$. 
We discretize the domain using a $50\times 50$ spatial grid and 30 temporal collocation points. A hardware-efficient ansatz with CNOT entanglement is employed, the same the embedding network, training loop and optimizer are used as in the 1D case.
The results are visualized over five time steps in Fig.~\ref{fig:2d_reference} and Fig.~\ref{fig:2d_prediction}. The QPINN model accurately captures the diffusion behavior of the exact solution, with symmetric heat propagation and predicted profiles closely matching reference snapshots. The trained quantum circuit architecture (Fig.\ref{fig:2d_circuit}) consists of 6 qubits and 4 variational layers. Training dynamics (Fig.\ref{fig:2d_loss}) show rapid convergence of both the loss and mean squared error in early epochs.
\begin{figure}[htbp]
    \centering
    \begin{subfigure}[b]{0.48\textwidth}
        \centering
        \includegraphics[width=\textwidth]{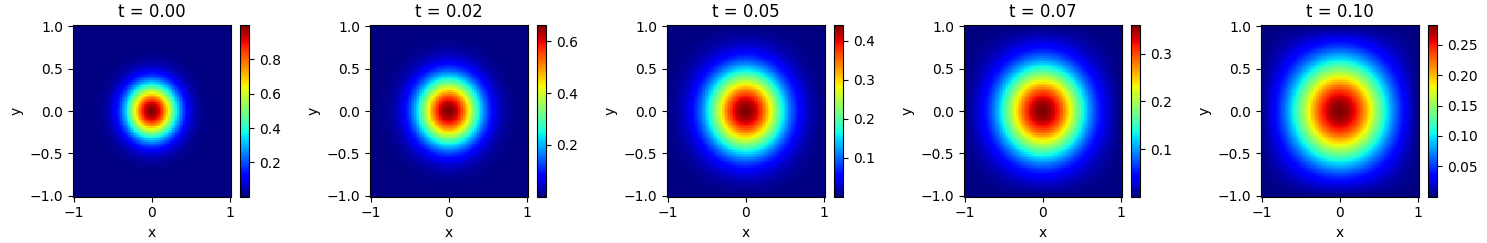}
        \caption{Reference solution of 2D heat equation}
        \label{fig:2d_reference}
    \end{subfigure}
    \hfill
    \begin{subfigure}[b]{0.48\textwidth}
        \centering
        \includegraphics[width=\textwidth]{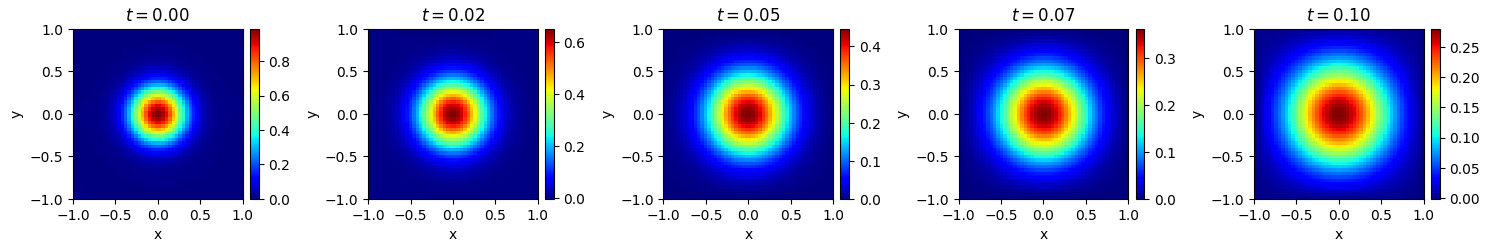}
        \caption{QPINN prediction for 2D heat equation}
        \label{fig:2d_prediction}
    \end{subfigure}
    \caption{\small Comparison of the 2D heat equation solution at selected time snapshots: (a) Reference solution of the 2D heat equation with Gaussian initial condition at corresponding time snapshots. (b) Predicted solution of the 2D heat equation using QPINNs.}
    \label{fig:2d_comparison}
\end{figure}
\vspace{-0.3cm}
\begin{figure}[htbp]
\centering
\includegraphics[width=0.45\textwidth]{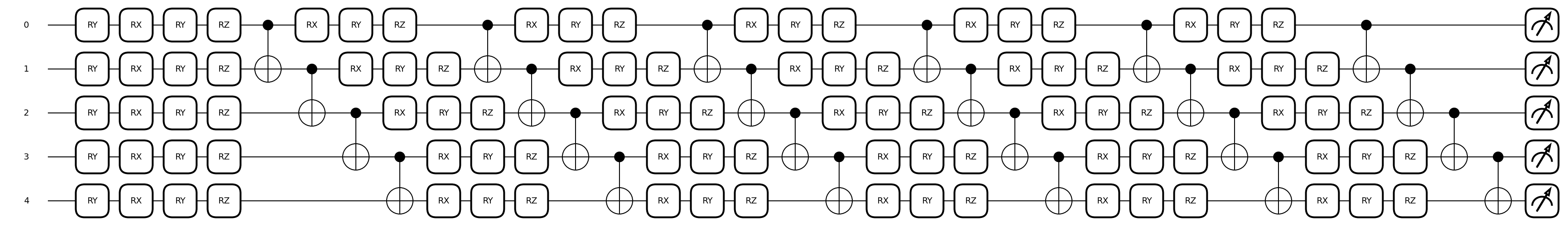}
\caption{\small Trained quantum circuit for the 2D QPINN simulation.}
\label{fig:2d_circuit}
\vspace{-0.5cm}
\end{figure}
\begin{figure}[htbp]
\centering
\includegraphics[width=0.45\textwidth]{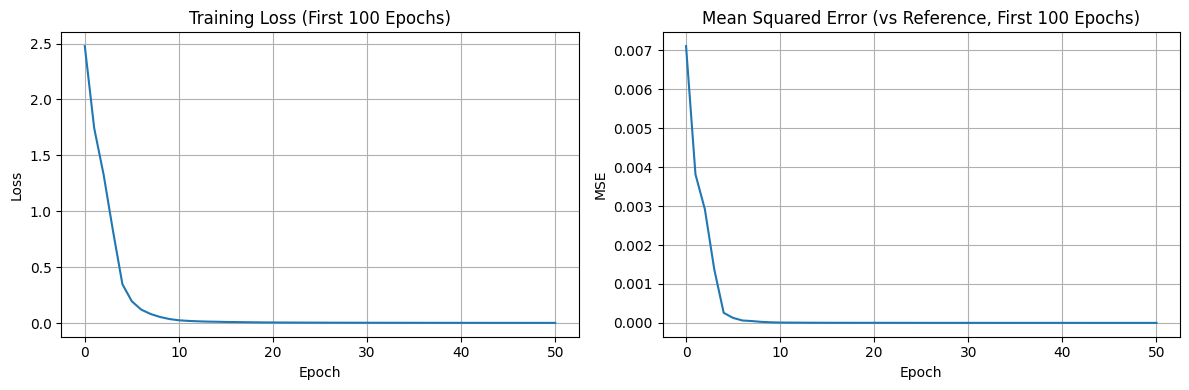}
\caption{\small Training loss and MSE convergence for the 2D heat equation.}
\label{fig:2d_loss}
\end{figure}


\vspace{-0.5cm}
\begin{thebibliography}{00}
\bibitem{b3}
N. B. Dehaghani, A. P. Aguiar, and R. Wisniewski, ``A Hybrid Quantum-Classical Physics-Informed Neural Network Architecture for Solving Quantum Optimal Control Problems,'' IEEE International Conference on Quantum Computing and Engineering (QCE), vol. 1, pp. 1378--1386, 2024.
\bibitem{b1} S. Berger, N. Hosters, and M. Möller, ``Trainable embedding quantum physics informed neural networks for solving nonlinear PDEs,'' Scientific Reports, vol. 15, pp. 1--14, 2025.
\bibitem{b2} O. Kyriienko, A. E. Paine, and V. E. Elfving, ``Solving nonlinear differential equations with differentiable quantum circuits,'' Physical Review A, vol. 103, pp. 052416, 2021.
\end{thebibliography}
\end{document}